\documentclass[epsfig,12pt]{article}

\usepackage[dvipdfm]{graphicx}
\usepackage{amssymb}
\usepackage{amsmath}
\usepackage{epsfig}
\usepackage{dcolumn}
\usepackage{rotating}
\usepackage{color}
\definecolor{rosso}{cmyk}{0,1,1,0.4}
\definecolor{rossos}{cmyk}{0,1,1,0.55}
\definecolor{rossoc}{cmyk}{0,1,1,0.2}
\definecolor{blu}{cmyk}{1,1,0,0.3}
\definecolor{blus}{cmyk}{1,1,0,0.6}
\definecolor{bluc}{cmyk}{1,1,0,0.1}
\definecolor{verde}{cmyk}{0.92,0,0.59,0.25}
\definecolor{verdec}{cmyk}{0.92,0,0.59,0.15}
\definecolor{verdes}{cmyk}{0.92,0,0.59,0.7}

\def\thalf{{\textstyle{\frac{1}{2}}}}
\def\tthalf{{\textstyle{\frac{3}{2}}}}

\newcommand{\be}{\begin{equation}}
\newcommand{\ee}{\end{equation}}
\newcommand{\ba}{\begin{eqnarray}}
\newcommand{\ea}{\end{eqnarray}}
\newcommand{\bd}{\begin{displaymath}}
\newcommand{\ed}{\end{displaymath}}

\newcommand{\bi}{\begin{itemize}}
\newcommand{\ei}{\end{itemize}}
\newcommand{\al}{\alpha}

\newcommand{\da}{\delta}
\newcommand{\la}{\lambda}

\newcommand{\oa}{\omega}

\newcommand{\cO}{{\cal O}}

\newcommand{\x}{\star}

\newcommand{\ra}{\rightarrow}
\newcommand{\Ra}{\Rightarrow}

\newcommand{\LT}{\left[}
\newcommand{\RT}{\right]}

\begin{document}
\tolerance=100000
\thispagestyle{empty}
\vspace{1cm}

\title{{\bf Thermodynamics of String Field Theory Motivated Nonlocal Models}}

\author{Tirthabir Biswas$^{1,2}$, Joseph Kapusta$^2$, Abraham Reddy$^2$}

\parindent=20pt

\maketitle

\begin{center}
{\it $^1$Department of Physics, Loyola University, New Orleans, LA 70118}\\
{\it $^2$School of Physics and Astronomy,
University of Minnesota, Minneapolis, MN 55455}
\end{center}

\vspace{1cm}

\begin{abstract}
We investigate the thermodynamic properties of the nonlocal tachyon motivated by their nonlocal structure in string field theory.  We use previously developed perturbative methods for nonlocal fields to calculate the partition function and the equation of state in the high temperature limit.  
%We find that in these models the tachyon condensation can be seen as a second order phase transition. 
We find that in these models the tachyons undergo a second order phase transition.  We compare our results with those of ordinary scalar field theory.  We also calculate the one loop finite temperature effective potential.

\end{abstract}

\newpage

%\setcounter{page}{1}

%\tableofcontents

\section{Introduction}
\setcounter{equation}{0}

Tachyons are ubiquitous in string theory. They made their first appearance in closed bosonic string theory. Since then, they have been found to exist as open string  excitations in the world volume of D-branes in bosonic theories and in non-BPS D-branes in superstring theory, as well as excitations in brane--anti-brane systems; see \cite{sen} for a review. In conventional field theory the appearance of a tachyon usually implies that we are perturbing around an unstable vacuum; the tachyon should evolve to its true vacuum with positive mass-squared around which we can perform perturbative quantum calculations meaningfully. There is growing evidence to suggest the same is true in string theory \cite{sen,sen2}. For instance, the open string tachyons are simply thought to represent the instability in the various D-brane systems.  The rolling of the tachyon from the unstable potential hill to its stable minimum then describes the dissolution of the unstable D-branes into closed string excitations around the true vacuum which no longer supports open string excitations. This process is often referred to as tachyon condensation in the string literature. While the classical dynamics of the above process has been studied extensively, relatively little attention have been paid to the quantum theory of the tachyon.

In the present paper, we will study thermodynamic properties of a class of nonlocal actions which, at least superficially, resembles the nonlocal action for the string field theory tachyons. To be more precise, the action we will consider is what one obtains in the simplest level of truncation  in the String Field Theory (SFT) approach \cite{witten} where only the tachyon field is kept \cite{Kostelecky:1989nt}. 
Our main interest will be to study whether the tachyon is stable above some critical temperature, which would imply that the brane configuration is stable.  If this is the case, then we have the possibility to study how the branes become unstable when the temperature drops below that critical value, leading to eventual dissolution of the branes.  Our analysis will demonstrate that, just as in ordinary field theory, one can consistently perform quantum calculations of the partition function in such nonlocal models to address these type of questions.

We would like to emphasize that while work done on such simplified level-truncated string-inspired nonlocal models have been fruitful in elucidating certain aspects of string theory %\cite{condensation,SS3,SS4,K,ZS,PS2} 
\cite{condensation}-\cite{PS2}, when considering temperatures above the string scale one expects all the string states to contribute to the partition function which our analysis does not account for\footnote{Our results for temperatures below the string scale should still provide insights into thermal SFT.}.  In particular, this means that we will be unable to capture any physics related to the conjectured stringy Hagedorn phase which is due the exponential growth of the spectrum of physical string states. Also, in the real string theory the mass of the tachyon is of the same order of magnitude as the string scale. For phenomenological reasons we will keep the mass of the tachyon arbitrary throughout the paper and, due to technicalities, we have only performed our calculations when the tachyon is much lighter than the string scale. For all these reasons the analysis and results presented in this paper may be of limited direct relevance to understanding the thermal properties of the complete SFT. However, we believe that the formalism and the computational techniques we have developed will help us to consider realistic stringy models in the future. With regard to understanding quantum phenomenon in string theory it is worth noting that similar nonlocal models, such as  p-adic strings \cite{PS1},  have been shown to reproduce properties such as thermal-duality \cite{PA} (which has been variously argued in the string literature \cite{atick}) and Regge behavior \cite{Minahan:2001wh}-\cite{TL1}.

To understand the origin of the nonlocal structure we will consider in our models, let us look at the SFT action.  Schematically, the string field $\Psi$ can be thought of as a matrix-valued 1-form \cite{Kostelecky:1989nt} with a Chern-Simons type action.  The bosonic SFT action, for instance, is then given by
\be
S={1\over 2\al'}\int \Psi\x Q\Psi+{g\over 3}\int \Psi\star\Psi\x\Psi
\label{bosonicSFT}
\ee
where $Q$ is the BRST operator which is normalized to provide canonical kinetic terms to the various particle fields contained in the SFT spectrum, $g$ is the string coupling constant, and $1/\sqrt{\al'}$ is the string tension.  The $\x$ product has the effect of diffusing or smearing out the interaction over the string length. For instance, if $f$ corresponds to a canonically normalized particle field in the SFT spectrum,  then an  interaction term  involving the $\x$ product in the string field $\Psi$ translates into interactions for $f$ where they only enter in momentum-dependent combinations
\bd
\tilde{f}=\exp\left[\al'\ln(3\sqrt{3}/4)\Box\right]f
\ed
Equivalently, one can work with the redefined fields $\tilde{f}$ in terms of which the interactions have the usual polynomial form, but the kinetic operator picks up the nonlocal exponential derivative dependence \cite{Kostelecky:1989nt}. Thus, if we keep only the tachyon field, then the corresponding field theory action has the form
\be
S=\int d^4x\ \LT \thalf \phi \, {\rm e}^{-{\Box/ M^2}}(\Box -m^2)\phi -V(\phi)\RT
\label{action}
\ee
where $m^2$ is the mass  of the tachyon at the maximum, and M is the scale of nonlocality that describes stringy interactions. Both $m$ and $M$ are proportional to the string tension.  For example, for the bosonic SFT $m^2=-1/\al'$, while $M^2\sim 1/\al'\ln(3\sqrt{3}/4)$.   We will treat $m$ and $M$ as independent parameters for technical and phenomenological reasons.  The $V(\phi)$ represents a polynomial interaction, typically cubic or quartic.

In this paper we employ finite temperature methods that were developed for such nonlocal theories in \cite{TL2,PA} to investigate the thermodynamic properties of the tachyonic excitations.  We will discover that, just as in ordinary Quantum Field Theory (QFT), the nolocal  tachyon undergoes a second order phase transition. The effect of the stringy nonlocality seems to weaken the phase transition. This means that the discontinuity in the specific heat as one approaches the critical temperature from above and below decreases as $M$ decreases. We emphasize that the QFT limit is expected to be attained in the limit $M \ra \infty$, and we explicitly verify that this is indeed the case.  In this paper we work in the limit $M\gg m$; that is, we are close to the particle limit, but the formalism and techniques that we have developed can be employed to understand the (perhaps) more interesting and relevant situation where $M \sim m$.  This will be reported in a future paper.

Another important motivation for considering such theories is to explore phenomenological applications to particle physics.  For instance,
our calculations help to clarify the relation between the conventional QFT, which is based on a renormalization prescription, and the string inspired nonlocal actions where loop diagrams are typically finite. Intuitively, the exponential cut-off scale, $M^2$, acts as a Lorentz-invariant physical regulator. The expressions for the various thermodynamic quantities in these nonlocal models are almost identical to what one  obtains using traditional renormalization prescription, except that there are corrections which are suppressed as  $\cO[\exp(-M^2/4T^2)]$. This happens because one takes the limit $M\ra\infty$ {\it after} imposing the ``renormalization conditions'', according to the standard renormalization prescription, whereas the $M$ in stringy Lagrangians is a finite physical parameter encoding the nonlocality of the model. This opens up possibilities for phenomenological applications in particle physics if $M$ is close enough to the scale of Standard Model physics.  For one proposed alternative to the Standard Model using SFT-type actions see~\cite{moffat}.

Another goal of this paper is also to pave the way for possible  connections between string theory and cosmology. For previous applications of nonlocal models to
%cosmology \cite{C2,C1,C3,SFTA,COSMO1,COSMO2,COSMO3}
cosmology, please see~\cite{C2}-\cite{COSMO3}.  (For related work on nonlocal gravity see
%\cite{moffat,GRAV1,GRAV2,gravity,khoury,vernov,moffat-gravity,modesto,odintsov}).
\cite{moffat}-\cite{odintsov}.) In recent years, string thermodynamics has found several applications in the early Universe cosmology \cite{thermal-cosmology}.  In particular, there have been efforts to see whether stringy thermal fluctuations can play a role in the formation of the anisotropies in the Cosmic Microwave Background Radiation (CMBR). There are also the so-called warm inflationary scenarios where particles are continuously produced and which then thermalize and influence inflationary dynamics, both at the level of the background and the fluctuations \cite{warm}. It would be interesting to consider similar scenarios where stringy excitations are produced instead, potentially providing a prospect to observe stringy properties in CMBR. Our calculations would be relevant for such a discussion.

The paper is organized as follows: In section 2, we introduce the nonlocal  model and its finite temperature formulation. By calculating the temperature dependence of the effective mass and minimum of the potential, we demonstrate that a second order phase transition occurs in the nonlocal models under consideration at high temperature, and estimate the critical temperature.  In section 3, we follow the traditional perturbation theory approach to calculate the 1 and 2-loop diagrams contributing to the partition function of the tachyon. This enables us to obtain the equation of state for the thermal excitations of the tachyons around the true minimum.  The same techniques can be used to calculate N-point diagrams for arbitrary values of N at high temperature. This is sufficient for us to obtain the critical temperature and determine the nature of the phase transition. In section 4, we summarize the analytical computations and provide numerical results.  This enables us to compare the equation of state in the nonlocal models with the analogous local QFT equation of state.  In section 5, we calculate the 1-loop effective potential which extrapolates away from the equilibrium states.  Concluding remarks are contained in section 6.

%%%%%%%%%%%%%%%%%%%%%%%%%%%%%%%%%%%%%%%%%%%%%%%%%
\section{Action and Critical Temperature}
\setcounter{equation}{0}

In this section we introduce the nonlocal  action and show that a second order phase transition is to be expected at high temperature.  More elaborate calculations of the equation of state and of the effective potential follow in sections 3, 4 and 5.

\subsection{Action}
Our starting point is the SFT type action given by \cite{SFTA}
\be
S=\int d^4x\ \LT \thalf \phi \, {\rm e}^{-{\Box/ M^2}}(\Box -m^2)\phi -V(\phi)\RT
\label{action}
\ee
The metric is such that $\Box = -\partial_t^2 + \nabla^2$.  In the bosonic cubic string field theory the action for the tachyon is of the above form with $V(\phi)\propto \phi^3$, while for the supersymmetric case one expects a quartic potential. In either case the mass-squared term is always negative, $m^2<0$, indicating the presence of a tachyon at $\phi=0$. In this paper we focus on the quartic coupling
\be
V(\phi)={\la}\phi^4
\ee
In SFT $\lambda$ is proportional to $g^2$.  Since the potential is bounded from below,  we expect to be able to perform loop calculations without encountering any pathologies. In this context we note that although the presence of higher derivative terms usually indicate the existence of additional ghost-like states, the fact that in the SFT action they combine into an exponential ensures that there are no extra poles in the propagator. Thus there are, in fact,  no additional physical states, ghosts or otherwise. There are also strong arguments to suggest that the initial value problem in these models are well defined, and classical trajectories can be uniquely specified by only a finite set of parameters \cite{ivp}.

There are two interesting limits of the action expressed by eq. (\ref{action}).  When $m$ is fixed and $M \rightarrow \infty$, one recovers the conventional local field theory action for a scalar field.  When $M$ is fixed and $m \rightarrow \infty$, one recovers the $p$-adic field theory \cite{PS1}.  In this paper we consider the case when both $m$ and $M$ are finite but with $|m| \ll M$.

Application of the usual finite temperature formalism to nonlocal actions involving an infinite series of higher derivative terms, such as (\ref{action}),  have been studied recently \cite{PA}. The basic prescription is rather straightforward and resembles the finite temperature methods implemented in usual local quantum field theories. The main difference is that the propagator gets modified by the presence of the nonlocal terms, and ultraviolet divergences in the quantum loops are either softened or eliminated altogether.   In some sense the mass parameter $M$ acts as an ultraviolet cutoff.  Thus the renormalization prescription is somewhat modified from the usual field theories.

One of the important differences between SFT-type theories and $p$-adic theory is that, in the latter case, $\phi=0$ corresponds to a minimum while in the SFT-type case it is a maximum, hence perturbative calculations around $\phi=0$ are not well-defined. (This is elaborated on in more detail in the following sections.)  To perform quantum loop calculations we must expand around the true minimum. At the classical, or tree, level the minimum is located at $\phi_0= \mu/ 2\sqrt{\la}$, where we have defined $\mu^2 \equiv -m^2>0$. See Fig. 1.
However, at finite temperature the minimum is shifted to smaller values of $\phi$. To account for this we expand around the true minimum, $v(T)$:
\be
\phi=v(T) +\phi_f
\ee
where $v(T)$ is independent of space and time but does depend on the temperature, while $\phi_f$ is the fluctuation around it whose average value is zero.
\begin{figure}[!htbp]
\begin{center}
\includegraphics[width=4.0in]{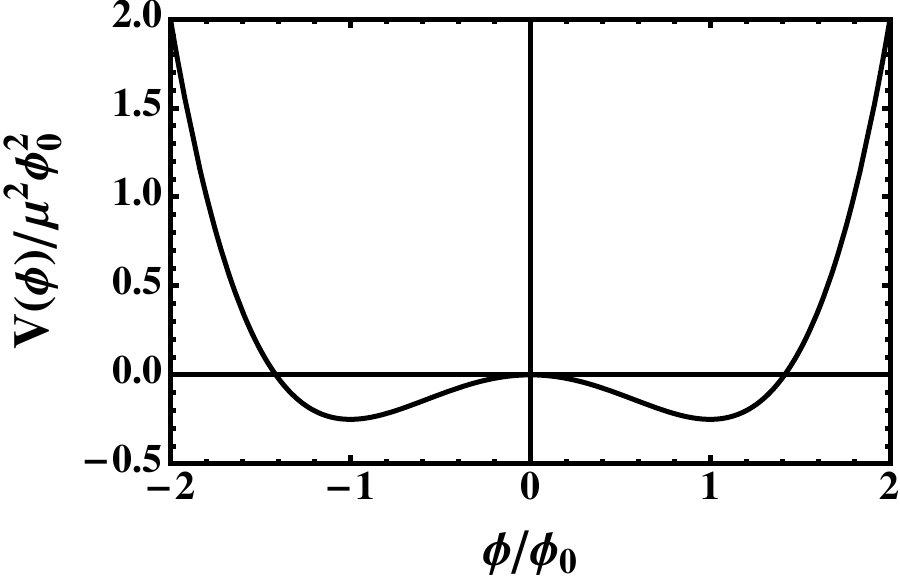}
\end{center}
\caption{ The vacuum potential $V$ in dimensionless form.}
\end{figure}
When doing loop calculations using the fluctuation $\phi_f$ around the average value of the field $v(T)$ at the temperature $T$, we should use the dressed propagator (or our best estimate of it) and not the bare propagator.  As in conventional models of spontaneously broken symmetries in local field theory, we will see that as the temperature increases the values of the condensate $v(T)$ and the effective mass both decrease to zero at a critical temperature $T_c$.  This can be done in the usual way by introducing an additional mass-shift $\delta m^2$ in the the quadratic part of the action and then subtracting it as a counter-term. Then the Lagrangian reads
\be
{\cal L} = {\cal L}_{\rm quad} + {\cal L}_{\rm int} + {\cal L}_{\rm ct} - V_{\rm cl}(v)
\ee
where
\ba
{\cal L}_{\rm quad} &=& \thalf \phi_f \left[ {\rm e}^{-\Box/M^2} \left( \Box + \mu^2 \right) -12\lambda v^2 - \delta m^2 \right] \phi_f
\nonumber \\
{\cal L}_{\rm int} &=& -4\lambda v \phi_f^3 - \lambda \phi_f^4
\nonumber \\
{\cal L}_{\rm ct} &=& \thalf \delta m^2 \phi_f^2
\nonumber \\
V_{\rm cl}(v) &=& - \thalf \mu^2 v^2 + \lambda v^4
\label{xi-independent} \\
\ea
The $v(T)$ and $\da m^2(T)$ are as yet undetermined functions of the temperature.

%%%%%%%%%%%%%%%%%%%%%%%%%%
\subsection{Critical Temperature}

The functions $v(T)$ and $\da m^2(T)$ introduced above can be determined by the 1-point and 2-point loop diagrams, respectively.   In this section we derive formulas for them at the 1-loop mean-field level.  We begin by noting that the thermodynamic potential is given by
\be
\Omega = V_{\rm cl}(v)  - \frac{T}{V} \ln \left\{ \int [d\phi_f]
\exp \left( \int_0^{\beta} d\tau \int_V d^3x \left[ {\cal L}_{\rm quad}+{\cal L}_{\rm int}+{\cal L}_{\rm ct} \right] \right) \right\}
\ee
where $\beta=1/T$ and $V$ is the volume.  The thermodynamic potential must be a minimum with respect to variations in $v$, thus providing an equation to determine $v(T)$.  Similarly the function $\delta m^2$ is determined by the Schwinger-Dyson equation in the 1-loop mean-field approximation.

According to the formalism that was developed in \cite{PA} to compute Feynman diagrams in nonlocal field theories such as (\ref{action}), the only diagrammatic rule that needs to change is the propagator. From (\ref{xi-independent}) we see that
\be
{\cal D} = {1\over {\rm e}^{(\omega_n^2 + p^2)/M^2} \left( \omega_n^2 + p^2 - \mu^2 \right) + 12 \lambda v^2 + \delta m^2}
\label{dressed}
\ee
where we have introduced the Matsubara frequencies, $\oa_n=2n\pi T$, as appropriate for finite temperature field theory \cite{FT}. We emphasize the exponential suppression of the propagator at large momenta which is a typical characteristic of these theories and which helps to regulate the ultarviolet divergences of loop diagrams.

Let us now try to compute these quantities at the 1-loop level; a more rigorous comprehensive analysis will be provided in the next two sections.
\be
\frac{\partial \Omega}{\partial v} =0 \;\;\; \Ra \;\;\;
(4\lambda v^2 -\mu^2)v + 12 \lambda v  T \sum_n \int \frac{d^3p}{(2\pi)^3} {\cal D}=0
\ee
where the first term comes from the classical potential while the second term comes from the 1-loop tadpole diagram.  For sufficiently small $T$ there is a local maximum at $v=0$ and a minimum at
\be
v^2 = \frac{\mu^2}{4\lambda} - 3 T \sum_n \int \frac{d^3p}{(2\pi)^3} {\cal D}(\omega_n,p;v,\delta m^2)
\label{condensate}
\ee
At a critical temperature $T_c$ these become degenerate.  The value of $T_c$ is determined by
\be
T_c =\frac{\mu^2}{12\lambda}\LT\sum_n \int \frac{d^3p}{(2\pi)^3} {\cal D}\RT^{-1}
\ee
so that above this temperature there is only one minimum at $v=0$. This temperature will later be identified with the critical temperature of a second order phase transition.

The two equations above are not sufficient to determine $v(T)$ and $T_c$ because the right side depends on ${\cal D}$ which itself depends on $\delta m^2$.  In the mean-field approximation it is determined by the 1-loop self-energy diagram which involves only the 4-point vertex, not the other diagram involving the 3-point vertices.  (For details see the next section.)  This is a gap equation.
\be
\delta m^2 = 12 \lambda T \sum_n \int \frac{d^3p}{(2\pi)^3} {\cal D}(\omega_n,p;v,\delta m^2)
\label{gapequation}
\ee
Equations (\ref{condensate}) and (\ref{gapequation}) are to be solved simultaneously and self-consistently to determine $v(T)$ and $\delta m^2(T)$.

First consider $T>T_c$: the minimum is located at $v=0$ and so $\da m^2$ can be determined as the solution to the single equation
\be
\delta m^2 = 12 \lambda T \sum_n \int \frac{d^3p}{(2\pi)^3} {1\over  {\rm e}^{(\omega_n^2 + p^2)/M^2} \left( \omega_n^2 + p^2 - \mu^2 \right) + \da m^2}
\label{deltam-aboveTc}
\ee

For $T<T_c$, on the other hand, there is a simple relation between $\da m^2$ and $v$, namely
\be
\delta m^2 = \mu^2 - 4 \lambda v^2
\ee
The temperature $T_c$ where $v$ goes to zero is exactly the same temperature where the effective mass
\be
m^2_{\rm eff} \equiv \delta m^2 - \mu^2 + 12\lambda v^2 = 8 \lambda v^2
\ee
goes to zero.  The condensate can be determined via
\be
v^2 = \frac{\mu^2}{4\lambda} - 3 T  \sum_n \int \frac{d^3p}{(2\pi)^3} {1\over  {\rm e}^{(\omega_n^2 + p^2)/M^2} \left( \omega_n^2 + p^2 - \mu^2 \right) + \mu^2 + 8 \lambda v^2}
\label{deltam-belowTc}
\ee
and this in turn allows for the direct algebraic determination of $\delta m^2$.

To obtain simple analytic results, for the moment we shall assume that $\mu \ll M$ and that $M \ll T_c$.  These assumptions can of course be relaxed albeit at the expense of numerical calculations, and they will be presented in the following sections.  Therefore we focus on temperatures $T \gg M$.  In this situation a nonzero Matsubara frequency will contribute an amount which is suppressed by a factor $\exp(-4\pi^2 T^2/M^2)$ and is totally ignorable.  Thus
\be
T \sum_n \int \frac{d^3p}{(2\pi)^3} {\cal D} \rightarrow \frac{MT}{4\pi\sqrt{\pi}} \, .
\ee
Hence both the condensate $v$ and the effective mass $m_{\rm eff}^2 = 12 \lambda v^2 - \mu^2 + \delta m^2$ must vanish at the same critical temperature given by
\be
T_c = \frac{\pi \sqrt{\pi}}{3} \frac{\mu^2}{\lambda M} \, .\
\ee
This is strongly indicative of a second order phase transition.  For consistency we need $T_c \gg M$.  This results in the limit
\be
\lambda M^2 \ll \mu^2 \ll M^2 \, .
\ee
To summarize: in this limit
\ba
v^2 &=& \frac{\mu^2}{4\lambda} - \frac{3MT}{4\pi\sqrt{\pi}} \nonumber \\
\delta m^2 &=& \frac{3\lambda MT}{\pi \sqrt{\pi}} \nonumber \\
m_{\rm eff}^2 &=& 2\mu^2 - \frac{6\lambda MT}{\pi \sqrt{\pi}} = 8\lambda v^2
\ea
below $T_c$ and
\ba
v^2 &=& 0 \nonumber \\
\delta m^2 &=& \frac{3\lambda MT}{\pi \sqrt{\pi}} \nonumber \\
m_{\rm eff}^2 &=& \frac{3\lambda MT}{\pi \sqrt{\pi}}-\mu^2
\label{m_at_highT}
\ea
above $T_c$.

%%%%%%%%%%%%%%%%%%%%%%%%%%%%%%

\section{Equation of State}
\setcounter{equation}{0}

In this section we perform a more sophisticated analysis of the equation of state.   We proceed analytically as far as possible and defer numerical calculations to a later section.  Readers primarily interested in the results may skip this section.

%%%%%%%%%%%%%%%%%%%%%%%

\subsection{Formalism}

The Lagrangian we will work with is
\be
{\cal L}_{\rm quad} = \thalf \phi \;{\rm e}^{-\Box/M^2} \left( \Box + \mu^2 \right) \phi + \thalf \gamma \phi^2 - \lambda \phi^4
\ee
Here we have added a counter-term $\thalf \gamma \phi^2$ with a coefficient $\gamma$ which will be adjusted so that the value of the condensate in the vacuum is the same as the classical expression $\mu^2/4\lambda$.  It also insures that no new poles are introduced into the propagator.  As before we represent the field in the form
\be
\phi = v(T)+\phi_f
\ee
where $v(T)$ is the equilibrium value of the condensate at temperature $T$ and $\phi_f$ is the fluctuation around it whose average value is zero.  After making this shift, and acknowledging that terms linear in $\phi_f$ will average to zero in the functional integral, the Lagrangian can be written as
\be
{\cal L} = {\cal L}_{\rm quad} + {\cal L}_{\rm int} + {\cal L}_{\rm ct} - V_{\rm cl}(v)
\ee
where
\ba
{\cal L}_{\rm quad} &=& \thalf \phi_f \left[ {\rm e}^{-\Box/M^2} \left( \Box + \mu^2 \right) -12\lambda v^2 - \delta m^2 \right] \phi_f
\nonumber \\
{\cal L}_{\rm int} &=&  - 4\lambda v \phi_f^3 - \lambda \phi_f^4
\nonumber \\
{\cal L}_{\rm ct} &=& \thalf \left(\delta m^2 + \gamma \right) \phi_f^2
\nonumber \\
V_{\rm cl}(v) &=& - \thalf \left( \mu^2 + \gamma \right) v^2 + \lambda v^4 \, .
\ea
An additional mass shift $\delta m^2$ has been added to the quadratic part of the action and then subtracted as a counter-term.  The reason is that as the temperature increases the value of the condensate $v$ decreases to zero at a critical $T_c$, and therefore at some temperature below $T_c$ the effective squared mass in the propagator becomes negative.  This just means that we should do our calculations with the dressed propagator (or our best estimate of it) and not the mean field propagator.  The value of $\delta m^2$ has to be determined at each temperature self-consistently just like $v$ does.  The $V_{\rm cl}(v)$ is the contribution to the classical potential from the condensate field.

The Feynman rules corresponding to the above action are as follows.  The dressed propagator ${\cal D}$ in the imaginary time formalism is given by
\be
{\cal D}^{-1} = {\rm e}^{(\omega_n^2 + p^2)/M^2} \left( \omega_n^2 + p^2 - \mu^2 \right) + 12 \lambda v^2 + \delta m^2
\ee
while the mean field propagator is given by
\be
\bar{\cal D}^{-1} = {\rm e}^{(\omega_n^2 + p^2)/M^2} \left( \omega_n^2 + p^2 - \mu^2 \right) + 12 \lambda v^2 \, .
\ee
The thermodynamic potential is
\be
\Omega = V_{\rm cl}(v) - \frac{T}{V} \ln \left\{ \int [d\phi_f]
\exp \left( \int_0^{\beta} d\tau \int_V d^3x \left[ {\cal L}_{\rm quad}+{\cal L}_{\rm int}+{\cal L}_{\rm ct} \right] \right) \right\}
\ee
where $\beta=1/T$ and $V$ is the volume.  In a diagrammatic expansion the field $\phi_f$ is represented by a solid line.  The vertices can easily be read off from the expressions above.  The quartic interaction $\phi_f^4$ has the vertex $-\lambda$ and the cubic interaction $\phi_f^3$ has the vertex $-4\lambda v$.  A cross or X represents the counter-term $-(\delta m^2 + \gamma)$.

The thermodynamic potential can be considered a function of the equilibrium condensate and a functional of the dressed propagator \cite{LeeMargulies,FT}.
\be
\Omega = V_{\rm cl}(v) - \thalf T \sum_n \int \frac{d^3p}{(2\pi)^3} \left\{ \ln \left( T^2 {\cal D}\right) - {\cal D}\bar{\cal D}^{-1} + 1 \right\}
+ \sum_{l=2}^{\infty} \Omega_l(v,{\cal D})
\ee
The $\Omega_l$ is the $l$-loop contribution to the potential.  (A counter-term counts as one loop in this context.)  Extremizing with respect to $v$ removes tadpole diagrams, and extremizing with respect to ${\cal D}$ removes one particle irreducible diagrams.  In the same way one could remove two particle irreducible diagrams by introducing dressed vertices.  The equations that determine the equilibrium solution are
\be
\frac{\partial \Omega}{\partial v} = -(\mu^2 + \gamma)v + 4\lambda v^3 + 12 \lambda v T \sum_n \int \frac{d^3p}{(2\pi)^3} {\cal D}
+ \sum_{l=2}^{\infty} \frac{\partial \Omega_l(v,{\cal D})}{\partial v} = 0
\ee
and
\be
{\cal D}^{-1} - \bar{\cal D}^{-1} =2 \sum_{l=2}^{\infty} \frac{\delta \Omega_l(v,{\cal D})}{\delta {\cal D}}
\ee
the latter being the Schwinger-Dyson equation.  Due to the functional derivative the difference of the inverse propagators is frequency and momentum dependent in general.  Thus $\delta m^2$ should in principle be the self-energy $\Pi(\omega_n,p)$.  However, in the approximations used in this paper a constant $\delta m^2$ will suffice.  Terminating the expansion at two loops results in the diagrams displayed in Fig. 2.  We would like to point out that all loop diagrams are UV finite on account of the exponential damping of the propagator, and they are IR finite except at a second order critical temperature where the mass vanishes (correlation length diverges).

\begin{figure}[!htbp]
\begin{center}
\includegraphics{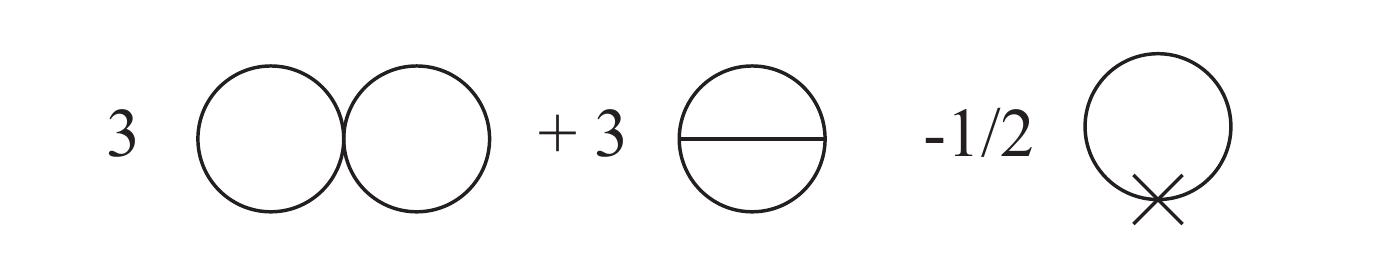}
\end{center}
\caption{ Two-loop contribution to $\ln Z$ including combinatoric factors and the counter-term.}
\end{figure}

\begin{figure}[!htbp]
\begin{center}
\includegraphics{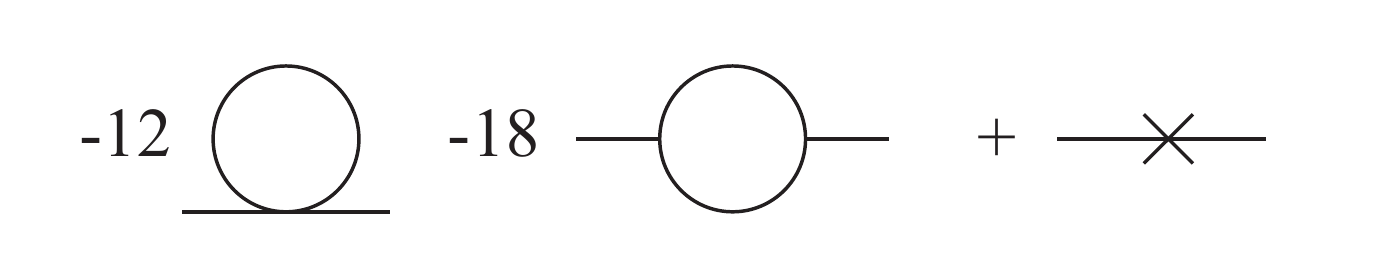}
\end{center}
\caption{One loop contribution to the self-energy including combinatoric factors and the counter-term.}
\end{figure}

The behavior we expect is that the zero temperature condensate decreases with increasing temperature due to thermal fluctuations.  If it goes to zero at a finite temperature $T_c$ then a phase transition ought to have occurred.  Now suppose that perturbation theory can be applied, at least if we are not too close to $T_c$.  To 1-loop order the equation for $v(T)$ is
\be
v^2 = \frac{\mu^2+\gamma}{4\lambda} - 3 T \sum_n \int \frac{d^3p}{(2\pi)^3} {\cal D} \, .
\ee
The mass shift $\delta m^2$ can also be calculated.  At 1-loop order it receives contributions from the diagrams shown in Fig. 3, which are obtained from those in Fig. 2.  We will neglect the diagram involving the three point vertex; this will be justified {\it a posteriori}.
\be
\delta m^2 = 12 \lambda T \sum_n \int \frac{d^3p}{(2\pi)^3} {\cal D} - \gamma
\ee
Therefore, to 1-loop order
\be
\delta m^2 = \mu^2 - 4 \lambda v^2 \, ,
\ee
and the propagator to this order is
\be
{\cal D}^{-1} = {\rm e}^{(\omega_n^2 + p^2)/M^2} \left( \omega_n^2 + p^2 - \mu^2 \right) + \mu^2 + 8 \lambda v^2 \, .
\ee
The temperature $T_c$ where $v$ goes to zero is the same temperature where the mass $\delta m^2 - \mu^2$ goes to zero.

The counter-term coefficient $\gamma$ can readily be determined by calculating $\delta m^2$ at $T=0$.  In the $T \rightarrow 0$ limit
\bd
T \sum_n \rightarrow \int \frac{dp_4}{2\pi}
\ed
Under the assumption that $\mu \ll M$ the integral is trivial, and the requirement that $\delta m^2 = 0$ requires
\be
\gamma = \frac{3\lambda M^2}{4\pi^2}
\ee
Working to higher order in the loop expansion would give the expansion of $\gamma$ in a power series in $\lambda$.

Now let us consider the contribution of the 2-loop contributions to the thermodynamic potential.  Referring to Fig. 2 they are
\be
\Omega_{\rm quartic} =3 \lambda \left[ T \sum_n \int \frac{d^3p}{(2\pi)^3} {\cal D}(\omega_n,p) \right]^2
\ee
and
\be
\Omega_{\rm cubic} = -48 \lambda^2 v^2  T \sum_{n_1} \int \frac{d^3p_1}{(2\pi)^3}
 T \sum_{n_2} \int \frac{d^3p_2}{(2\pi)^3} {\cal D}(\omega_{n_1},{\bf p}_1) {\cal D}(\omega_{n_2},{\bf p}_2)
 {\cal D}(\omega_{n_1 + n_2},{\bf p}_1+{\bf p}_2)
 \ee
in an obvious notation.  As emphasized before, the zero Matsubara mode dominates at high temperature, $T \gg M$, but still $T < T_c$ so that the cubic interaction does not vanish.  (If it did then clearly $\Omega_{\rm cubic}=0$.)  Hence
\be
\Omega_{\rm quartic} \sim \lambda M^2 T^2
\ee
and
\be
\Omega_{\rm cubic} \sim \lambda^2 v^2 T^2 \ln\left(M^2/\lambda v^2\right) \, .
\ee
The diagram with the cubic vertices is smaller by a factor of $\lambda v^2/M^2$ which tends to zero at $T_c$ even when including the logarithmic factor.

Similar conclusions can be reached for the self-energy.  Using the dressed propagator of eq. (\ref{dressed}) as reference we find that
\be
\Pi_{\rm quartic}(\omega_n,p) = 3\mu^2 - \frac{6\lambda MT}{\pi \sqrt{\pi}}
\ee
which is frequency and momentum independent and has the limit $2\mu^2$ at $T_c$ .  In contrast, at large frequency and momentum
\be
\Pi_{\rm cubic}(\omega_n,p) \sim \frac{\lambda^2 v^2 MT}{\omega_n^2 + p^2}
\exp\left[ - \left( \omega_n^2 + p^2\right)/M^2\right]
\ee
which is exponentially suppressed.  At zero frequency and momentum
\be
\Pi_{\rm cubic}(0,0) = - \frac{144 \lambda^2 T^2}{\pi^2} \int_0^{\infty} \frac{dk\,k^2}{(k^2+m_{\rm eff}^2)^2}
= - \frac{36\lambda^2 v^2 T}{\pi m_{\rm eff}} = -\frac{9\sqrt{2}}{\pi} \lambda^{3/2} v T \, .
\label{0limitPI}
\ee
which goes to zero at $T_c$.  This justifies the neglect of the diagrams involving the cubic interactions at high temperature.

Let us summarize the approximate expression for the thermodynamic potential.  It includes the two loop diagram involving the four point vertex but not the two loop diagram involving the three point vertex; as argued previously, the latter is suppressed when $\lambda \ll 1$ and $\mu \ll M$.
\bd
\Omega = -\thalf \left(\mu^2 + \gamma\right) v^2 + \lambda v^4 + \thalf T \sum_n \int \frac{d^3p}{(2\pi)^3} \ln \left( \beta^2 {\cal D}^{-1}\right)
\ed
\be
-\thalf \left( \delta m^2 + \gamma\right) T \sum_n \int \frac{d^3p}{(2\pi)^3} {\cal D} +3\lambda \left[ T \sum_n \int \frac{d^3p}{(2\pi)^3} {\cal D} \right]^2
\ee
The condensate satifies
\be
v^2 = \frac{\mu^2+\gamma}{4\lambda} - 3 T \sum_n \int \frac{d^3p}{(2\pi)^3} {\cal D}
\ee
below $T_c$ while $v=0$ above $T_c$.  The propagator is
\be
{\cal D}^{-1} = {\rm e}^{(\omega_n^2 + p^2)/M^2} \left( \omega_n^2 + p^2 - \mu^2 \right) + \mu^2 + 8 \lambda v^2
\ee
below $T_c$ and
\be
{\cal D}^{-1} = {\rm e}^{(\omega_n^2 + p^2)/M^2} \left( \omega_n^2 + p^2 - \mu^2 \right) + \delta m^2
\ee
above $T_c$, where
\be
\delta m^2 = 12 \lambda T \sum_n \int \frac{d^3p}{(2\pi)^3} {\cal D} - \gamma \, .
\ee
It is useful to express the propagator as
\be
{\cal D}^{-1}\left(\omega_n,p;m_{\rm eff}^2\right) = {\rm e}^{(\omega_n^2 + p^2)/M^2} \left( \omega_n^2 + p^2 - \mu^2 \right) + \mu^2 + m_{\rm eff}^2 \, .
\label{effectiveprop}
\ee
The quantity $m_{\rm eff}$ is approximately the pole mass when it is small compared to $M$.  Rigorously speaking it is the screening mass.  At zero temperature $m_{\rm eff}^2 = 2\mu^2$ and is always non-negative, vanishing only at $T_c$.  Below $T_c$ we must solve for $v(T)$ self-consistently, and the result then also determines $m_{\rm eff}(T)$.  Above $T_c$ we must solve for $\delta m^2(T)$ self-consistently, and this determines $m_{\rm eff}(T)$.

%%%%%%%%%%%%%%%%%%%%%%%%
\subsection{Sums and Integrals}

Let us calculate the relevant sums and integrals.  We are interested in temperatures $T > \mu$ (previously, in section 2,  we considered $T > M$) and mass scales $M > \mu$.  Let us start with the oft-appearing quantity
\bd
T \sum_n \int \frac{d^3p}{(2\pi)^3} {\cal D} \, .
\ed
Note that it is convergent in both the IR and UV.  To the desired order
\be
T \sum_n \int \frac{d^3p}{(2\pi)^3} {\cal D} \approx  T \sum_n \int \frac{d^3p}{(2\pi)^3} \frac{{\rm e}^{-(\omega_n^2+p^2)/M^2}}
{\omega_n^2+p^2}
\label{Dexpansion}
\ee
The trick is to use the integral representation
\be
\sum_n \frac{{\rm e}^{-(\omega_n^2+p^2)/M^2}}{\omega_n^2+p^2}=
\frac{1}{M^2}\sum_n \int_1^{\infty} d\alpha^2 {\rm e}^{-\alpha^2(\omega_n^2+p^2)/M^2}
\ee
and then use the function
\be
\zeta(s) = \sum_{n=-\infty}^{\infty} {\rm e}^{-s^2n^2}
\ee
which appears so often in the study of the $p$-adic theory at finite temperature \cite{PA}.  After integrating over momentum
\be
T \sum_n \int \frac{d^3p}{(2\pi)^3} \frac{{\rm e}^{-(\omega_n^2+p^2)/M^2}}{\omega_n^2+p^2} = \frac{MT}{4\pi\sqrt{\pi}}
f(T/M)
\ee
where we have defined the function
\be
f(T/M) \equiv \int_1^{\infty} \frac{d\alpha}{\alpha^2} \zeta\left(\frac{2\pi T}{M}\alpha\right)
\ee
Therefore, to the desired accuracy
\be
T \sum_n \int \frac{d^3p}{(2\pi)^3} {\cal D} = \frac{MT}{4\pi\sqrt{\pi}} f(T/M) \, .
\ee

The integral over $\alpha$ can be performed numerically, but it can also be calculated in the low and high temperature limits.  These calculations are facilitated by an interesting property of the $\zeta$ function that
\be
\zeta(s)=\frac{\sqrt{\pi}}{s} \zeta \left(\frac{\pi}{s}\right) \, .
\ee
When $s > \sqrt{\pi}$
\be
\zeta(s) = 1+2{\rm e}^{-s^2}+ 2{\rm e}^{-4s^2}+\cdot\cdot\cdot
\ee
and when $s < \sqrt{\pi}$
\be
\zeta(s) = \frac{\sqrt{\pi}}{s}\left( 1+2{\rm e}^{-\pi^2/s^2}+ 2{\rm e}^{-4\pi^2/s^2}+\cdot\cdot\cdot \right) \, .
\ee
Using the above, we can calculate the behavior for $T \le T_0$ and for $T \ge T_0$ where $T_0 \equiv M/2\sqrt{\pi}$.  For $T \ge T_0$
\be
f(T/M) = 1 + 2\sum_{n=1}^{\infty} \left\{ {\rm e}^{-4\pi^2n^2T^2/M^2}
- \frac{2\pi\sqrt{\pi}nT}{M} \left[ 1 - \Phi\left(\frac{2\pi nT}{M}\right) \right] \right\}
\ee
where $\Phi$ is the probability integral
\be
\Phi(u) = \frac{2}{\sqrt{\pi}} \int_0^u dt \, {\rm e}^{-t^2} \, .
\ee
This has the high temperature expansion
\be
f(T/M) =
1 + \frac{M^2}{4\pi^2 T^2} {\rm e}^{-4\pi^2T^2/M^2} + {\cal O}\left( \frac{M^4}{T^4} {\rm e}^{-4\pi^2T^2/M^2} \right) \, .
\ee

For $T \le T_0$ the integral can be broken up into two pieces, one from $\alpha = 1$ to $\alpha = \alpha_0$ and another from $\alpha = \alpha_0$ to $\alpha = \infty$, where $\alpha_0 = M/2\sqrt{\pi}T$.
\bd
f(T/M) = \frac{2\sqrt{\pi}T}{M} + \frac{M}{4\sqrt{\pi}T} \left( 1 - \frac{4\pi T^2}{M^2} \right)
\ed
\be
+ \frac{4\sqrt{\pi}T}{M} \sum_{n=1}^{\infty} \left[ \frac{1}{2\pi n^2} \left( {\rm e}^{-n^2\pi} - {\rm e}^{-n^2M^2/4T^2} \right)
+ {\rm e}^{-n^2\pi} - n\pi \left( 1 - \Phi(n\sqrt{\pi}) \right) \right]  \, .
\ee
The low temperature limit is
\be
f(T/M) = \frac{M}{4\sqrt{\pi}T} + \frac{\pi \sqrt{\pi}}{3} \frac{T}{M}
+ {\cal O}\left(\frac{T}{M} {\rm e}^{-M^2/4T^2}\right)
\ee
where we have used
\be
1 + 2\sum_{n=1}^{\infty} \left[ \left( \frac{1}{\pi n^2} +2\right) {\rm e}^{-n^2\pi} -2\pi n \left( 1 - \Phi(n\sqrt{\pi}) \right) \right]  = \frac{\pi}{3} = 1.07163...
\ee
We have not proven the equality stated above, nor have we found reference to it in the literature, but it is true to any numerical accuracy that we have done.

Finally let us turn our attention to the one loop contribution
\bd
\thalf T \sum_n \int \frac{d^3p}{(2\pi)^3} \ln \left( \beta^2 {\cal D}^{-1}\right) \, .
\ed
It can be expressed as
\bd
\thalf T \sum_n \int \frac{d^3p}{(2\pi)^3} \left\{ \int_0^{m_{\rm eff}^2} \frac{d\alpha^2}{{\rm e}^{(\omega_n^2+p^2)/M^2}
\left( \omega_n^2 + p^2 -\mu^2\right) + \mu^2 + \alpha^2 }\right\}
\ed
\bd
+ \thalf T \sum_n \int \frac{d^3p}{(2\pi)^3} \ln\left[ \beta^2 \left({\rm e}^{(\omega_n^2+p^2)/M^2}
\left( \omega_n^2 + p^2 -\mu^2\right) + \mu^2 \right)\right] \, .
\ed
In the limit of small $\mu$ in comparison to $M$ and $T$ the second term can be written as
\bd
\thalf T \sum_n \int \frac{d^3p}{(2\pi)^3} \left\{  \frac{ \omega_n^2+p^2}{M^2} +\ln\left[ \beta^2 (\omega_n^2+p^2) \right]
- \frac{\mu^2 \left[ 1-{\rm e}^{-(\omega_n^2+p^2)/M^2)} \right]}{\omega_n^2+p^2} \right\} + {\cal O}(\mu^4)
\ed
The first term in curly brackets appears in the $p$-adic limit and it is zero \cite{PA}.  The second term contributes one massless bosonic degree of freedom.
\be
\thalf T \sum_n \int \frac{d^3p}{(2\pi)^3} \ln\left[ \beta^2 (\omega_n^2+p^2) \right] = -\frac{\pi^2}{90}T^4 + {\rm vacuum}
\ee
The third term can be written as
\bd
-\thalf \mu^2 \int \frac{d^3p}{(2\pi)^3} \frac{1}{p} \frac{1}{{\rm e}^{\beta p}-1}
-\thalf \mu^2 \int \frac{d^4p}{(2\pi)^4} \frac{1}{p^2}+ \frac{\mu^2 MT}{8\pi \sqrt{\pi}}f(T/M)
\ed
\be
= -\frac{ \mu^2 T^2}{24} + \frac{\mu^2 MT}{8\pi \sqrt{\pi}}f(T/M) + {\rm vacuum} \, .
\ee
See \cite{FT} for the integrals.  Hence, to the desired order
\be
\thalf T \sum_n \int \frac{d^3p}{(2\pi)^3} \ln \left( \beta^2 {\cal D}^{-1}(\omega_n,p;m_{\rm eff})\right)=-\frac{\pi^2}{90}T^4
-\frac{ \mu^2 T^2}{24}+\frac{m_{\rm eff}^2 + \mu^2}{8\pi \sqrt{\pi}} MTf(T/M) \, .
\ee

%%%%%%%%%%%%%%%%%%
\subsection{Equation of State for $M, T \gg \mu$}

Now we assemble what we have learned in the limit that $M \gg \mu$ and $T \gg \mu$.  The equation of state is expressed as pressure $P(T) = -\Omega(T)$ as a function of temperature $T$.  The pressure is normalized to zero at zero temperature.

For $T \le T_c$ the effective mass and condensate are given as functions of temperature by
\be
m_{\rm eff}^2 = 8\lambda v^2 = 2\mu^2 - \frac{3\lambda M^2}{2\pi^2} \left[ 4 \sqrt{\pi} \frac{T}{M} f\left(\frac{T}{M}\right) - 1\right]
\ee
The pressure is
\be
P = \frac{\pi^2}{90}T^4 + \frac{\mu^2 T^2}{24} -\frac{3\mu^2 M^2}{32\pi^2} \left[ 4 \sqrt{\pi} \frac{T}{M} f\left(\frac{T}{M}\right) - 1\right] + \frac{3\lambda M^4}{128\pi^4} \left[ 4 \sqrt{\pi} \frac{T}{M} f\left(\frac{T}{M}\right) - 1\right]^2
\ee
The entropy density $s(T)=dP(T)/dT$ can easily be computed by using the formula
\be
T\frac{df(T/M)}{dT} = f(T/M)-\zeta(2\pi T/M)
\ee
It is
\bd
s(T) = \frac{2\pi^2}{45} T^3 + \frac{\mu^2 T}{12}  +
\ed
\be
\left[ 2 f\left(\frac{T}{M}\right) - \zeta\left(\frac{2\pi T}{M}\right) \right] \left\{ \frac{3\lambda M^3}{16 \pi^3 \sqrt{\pi}} \left[ 4 \sqrt{\pi} \frac{T}{M} f\left(\frac{T}{M}\right) - 1 \right] - \frac{3\mu^2 M}{8 \pi \sqrt{\pi}} \right\}
\ee
The energy density is $\epsilon(T) = -P(T) + Ts(T)$.

For $T \ge T_c$ the condensate is zero and the effective mass is determined by the formula
\be
m_{\rm eff}^2 = \frac{3\lambda M^2}{4\pi^2} \left[4 \sqrt{\pi} \frac{T}{M} f\left(\frac{T}{M}\right) - 1  \right] - \mu^2
\ee
The pressure is
\be
P = \frac{\pi^2}{90}T^4 + \frac{\mu^2 T^2}{24} - \frac{3\lambda M^4}{256\pi^4}\left[ 4 \sqrt{\pi} \frac{T}{M} f\left(\frac{T}{M}\right) - 1\right]^2 -\frac{\mu^4}{16\lambda}
\ee
and the entropy density is
\be
s(T) = \frac{2\pi^2}{45} T^3 + \frac{\mu^2 T}{12}  - \frac{3\lambda M^3}{32 \pi^3 \sqrt{\pi}}\left[ 2 f\left(\frac{T}{M}\right) - \zeta\left(\frac{2\pi T}{M}\right) \right] \left[ 4 \sqrt{\pi} \frac{T}{M} f\left(\frac{T}{M}\right) - 1 \right]
\ee

Both the condensate and the effective mass vanish at the critical temperature $T_c$ determined by
\be
\frac{T_c}{M} f\left(\frac{T_c}{M}\right) = \frac{1}{4\sqrt{\pi}} + \frac{\pi\sqrt{\pi} \mu^2}{3\lambda M^2}
\ee
At this temperature both the pressure
\be
P(T_c)=\frac{\pi^2}{90}T_c^4 + \frac{\mu^2 T_c^2}{24}-\frac{\mu^4}{12\lambda}
\ee
and the entropy density
\be
s(T_c)=\frac{2\pi^2}{45} T_c^3 + \frac{\mu^2 T_c}{12}+\frac{\mu^2 M}{8\pi \sqrt{\pi}} \left[ \zeta\left(\frac{2\pi T_c}{M}\right) - 2 f\left(\frac{T_c}{M}\right) \right]
\ee
are continuous, but the heat capacity $c_V(T)=Tds(T)/dT$ is not.  Hence this is a second order phase transition.

%%%%%%%%%%%%%%%%%%%%%%%%%%%%%
\section{Numerical Results and Comparison to Local Field Theory}
\setcounter{equation}{0}

The string motivated field theory under study has three parameters: $M$, $\mu$ and $\lambda$.  What matters for the equation of state is not absolute magnitudes but relative magnitudes.  The perturbative analysis we have used requires that $\lambda \ll 1$.  However, since $\mu^2 > 0$ with the consequence of spontaneous symmetry breaking in the vacuum, the limit $\lambda = 0$ is not allowed since the theory would not be well-defined.  We have assumed that the string scale $M$ is large in comparison to the mass scale $\mu$.  We have also assumed that $T$ is large compared to $\mu$ but made no assumption about the ratio $T/M$.  From the point of view of conventional local field theory, $M$ acts as an ultraviolet regulator.  If one takes $M \rightarrow \infty$ in the action, one recovers the normal $\phi^4$ field theory with spontaneous symmetry breaking.  Let us examine the limits $T_c \ll M$ and $T_c \gg M$ analytically before turning to numerical calculations.

Suppose that $M \rightarrow \infty$ with $\mu^2/\lambda$ held fixed.  Then it is easy to show that
\be
m^2_{\rm eff}(T) = \left\{ \begin{array}{ll}
2\mu^2 \left( 1 - T^2/T_c^2 \right) & {\rm if} \; T \le T_c \\
\mu^2 \left( T^2/T_c^2 -1 \right) & {\rm if} \; T \ge T_c
\end{array} \right.
\ee
with $T_c^2 = \mu^2/\lambda$.  The condensate is determined by $8 \lambda v^2(T) = m^2_{\rm eff}(T)$ when $T \le T_c$ while $v(T)=0$ when $T \ge T_c$.  The equation of state below $T_c$ is
\ba
P(T) &=& \left( \frac{\pi^2}{90} + \frac{\lambda}{24} \right)T^4 -\frac{\mu^2T^2}{12} \nonumber \\
s(T) &=& 4\left( \frac{\pi^2}{90} + \frac{\lambda}{24} \right)T^3 -\frac{\mu^2T}{6} \nonumber \\
\epsilon(T) &=& 3\left( \frac{\pi^2}{90} + \frac{\lambda}{24} \right)T^4 -\frac{\mu^2T^2}{12} \nonumber \\
c_V(T) &=& 12\left( \frac{\pi^2}{90} + \frac{\lambda}{24} \right)T^3 -\frac{\mu^2T}{6}
\ea
and above $T_c$ is
\ba
P(T) &=& \left( \frac{\pi^2}{90} - \frac{\lambda}{48} \right)T^4 +\frac{\mu^2T^2}{24} -\frac{\mu^4}{16\lambda} \nonumber \\
s(T) &=& 4\left( \frac{\pi^2}{90} - \frac{\lambda}{48} \right)T^3 +\frac{\mu^2T}{12} \nonumber \\
\epsilon(T) &=& 3\left( \frac{\pi^2}{90} - \frac{\lambda}{48} \right)T^4 +\frac{\mu^2T^2}{24} +\frac{\mu^4}{16\lambda} \nonumber \\
c_V(T) &=& 12\left( \frac{\pi^2}{90} - \frac{\lambda}{48} \right)T^3 +\frac{\mu^2T}{12}
\ea
(Corrections to these formulas for large but finite $M$ are suppressed by the factor $\exp(-M^2/4T^2)$.)  Clearly $P$, $s$, and $\epsilon$ are continuous at $T_c$ but $c_V$ is not.  These are well-known, conventional finite temperature field theory results \cite{FT}.  Because $\lambda$ is required to be small, and $T \gg \mu$, the equation of state to first approximation is $\epsilon = 3P$.  To focus on the effect of interactions, especially near $T_c$, it is useful to define the dimensionless interation measure $(\epsilon - 3P)/\mu^2 T^2$.
From the above
\be
\frac{\epsilon-3P}{\mu^2T^2} = \left\{ \begin{array}{ll}
{\displaystyle \frac{1}{6}} & {\rm if} \; T \le T_c \\
{\displaystyle \frac{1}{4}\frac{T_c^2}{T^2} - \frac{1}{12}} & {\rm if} \; T \ge T_c
\end{array} \right.
\ee
Obviously it is continuous at $T_c$ but its derivative is not.

Now suppose that $\lambda M^2 \ll \mu^2$.  Then $T_c \gg M$ and so we focus on temperatures such that $T \gg M$.  In this case,
\be
m^2_{\rm eff}(T) = \left\{ \begin{array}{ll}
2\mu^2 \left( 1 - T/T_c \right) & {\rm if} \; T \le T_c \\
\mu^2 \left( T/T_c - 1 \right) & {\rm if} \; T \ge T_c
\end{array} \right.
\ee
with $T_c = \pi\sqrt{\pi}\mu^2/3\lambda M$.  The condensate is determined by $8 \lambda v^2(T) = m^2_{\rm eff}(T)$ when $T \le T_c$ while $v(T)=0$ when $T \ge T_c$.
The equation of state below $T_c$ is
\ba
P(T) &=& \frac{\pi^2}{90} T^4 +\left( \frac{\mu^2}{24} + \frac{3\lambda M^2}{8\pi^3} \right)T^2 - \frac{3\mu^2 MT}{8\pi \sqrt{\pi}} \nonumber \\
s(T) &=& 4 \frac{\pi^2}{90} T^3 +2\left( \frac{\mu^2}{24} + \frac{3\lambda M^2}{8\pi^3} \right)T - \frac{3\mu^2 M}{8\pi \sqrt{\pi}} \nonumber \\
\epsilon(T) &=& 3 \frac{\pi^2}{90} T^4 +\left( \frac{\mu^2}{24} + \frac{3\lambda M^2}{8\pi^3} \right)T^2 \nonumber \\
c_V(T) &=& 12 \frac{\pi^2}{90} T^3 +2\left( \frac{\mu^2}{24} + \frac{3\lambda M^2}{8\pi^3} \right)T
\ea
and above $T_c$ is
\ba
P(T) &=& \frac{\pi^2}{90} T^4 +\left( \frac{\mu^2}{24} - \frac{3\lambda M^2}{16\pi^3} \right)T^2 - \frac{\mu^4}{16\lambda} \nonumber \\
s(T) &=& 4\frac{\pi^2}{90} T^3 +2\left( \frac{\mu^2}{24} - \frac{3\lambda M^2}{16\pi^3} \right)T \nonumber \\
\epsilon(T) &=& 3\frac{\pi^2}{90} T^4 +\left( \frac{\mu^2}{24} - \frac{3\lambda M^2}{16\pi^3} \right)T^2 + \frac{\mu^4}{16\lambda} \nonumber \\
c_V(T) &=& 12\frac{\pi^2}{90} T^3 +2\left( \frac{\mu^2}{24} - \frac{3\lambda M^2}{16\pi^3} \right)T
\ea
(Corrections to these formulas are suppressed by the factor $\exp(-4\pi^2T^2/M^2)$.)  Once again, $P$, $s$, and $\epsilon$ are continuous at $T_c$ but $c_V$ is not.  The most noticeable difference is that in the conventional local field theory the effective mass-squared vanishes as $|T^2-T_c^2|$, whereas for the string field theory it vanishes as  $|T-T_c|$, although the difference is inconsequential in the limit $T \rightarrow T_c$.  The difference in the exponents is seen in the interaction measure too.
\be
\frac{\epsilon-3P}{\mu^2T^2} = \left\{ \begin{array}{ll}
{\displaystyle -\frac{1}{12}  + \frac{M}{8\pi \sqrt{\pi} T_c} \left( 9 \frac{T_c}{T} -2 \right)}& {\rm if} \; T \le T_c \\
{\displaystyle -\frac{1}{12} + \frac{M}{8\pi \sqrt{\pi} T_c} \left( 6 \frac{T_c^2}{T^2} + 1 \right)} & {\rm if} \; T \ge T_c
\end{array} \right.
\ee
Of course, our calculation is basically a mean-field approximation so the values of critical exponents cannot be taken as being very accurate.

It is instructive to examine the dependence of the discontinuity in the specific heat at the critical temperature as a function of $T_c/M$ to see the transition from conventional local field theory ($T_c/M \ll 1$) to the ``SFT" limit ($T_c/M \gg 1$).  In those two limits the discontinuity can be calculated analytically.
\be
c_V(T_c-)-c_V(T_c+) = \lambda T_c^3 \left\{ \begin{array}{ll}
{\displaystyle \frac{1}{2} }& {\rm if} \; T_c \ll M \\
{\displaystyle \frac{9}{8\pi^3} \left(\frac{M}{T_c}\right)^2} & {\rm if} \; T_c \gg M
\end{array} \right.
\ee
The discontinuity decreases monotonically with increasing $T_c/M$ when measured in units of $\lambda T_c^3$, the only sensible unit for comparison.

Now we show some full numerical results which do not make any assumption about the magnitude of $T_c/M$.  Figure 4 shows the dependence of $T_c/M$ on the variable $\mu^2/\lambda M^2$.  It changes from the square-root to linear dependence very rapidly when $T_c/M \sim 0.4$.  Figure 5 shows the dependence of $m_{\rm eff}^2/2\mu^2$ on $T/T_c$ for a value of $T_c/M \ll 1$ and for a value $T_c/M \gg 1$.  The figure clearly shows the quadratice dependence for small $T_c/M$ versus the linear dependence for large $T_c/M$.  Figure 6 shows the interaction measure, or deviation from the ideal relativistic equation of state $\epsilon = 3P$, for both small and large values of $T_c/M$.  Finally, Fig. 7 shows the discontinuity in the heat capacity at the critical temperature as a function of $T_c/M$; it decreases monontically towards zero as $T_c/M \rightarrow \infty$.

\begin{figure}[!htbp]
\begin{center}
\includegraphics[width=0.7\textwidth]{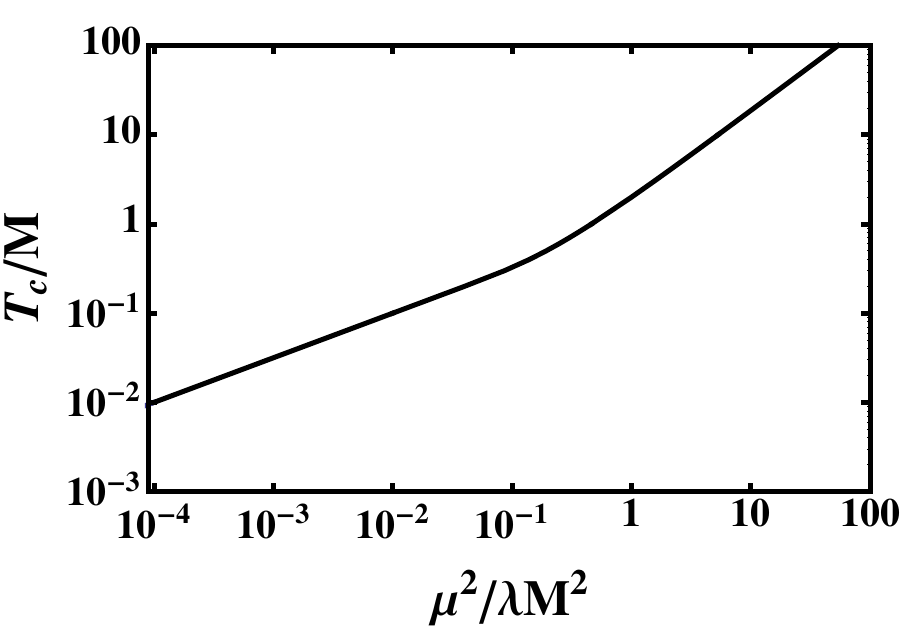}
\end{center}
\caption{Scaling of the critical temperature with the parameters.  The dependence changes from square-root to linear around $T_c/M \sim 0.4$.}
\end{figure}

\newpage

\begin{figure}[!htbp]
\begin{center}
\includegraphics[width=0.7\textwidth]{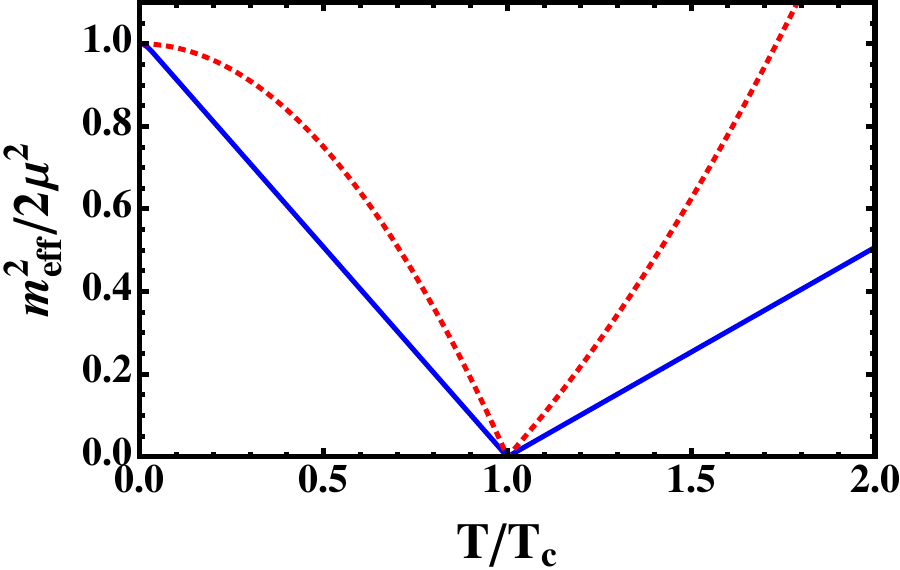}
\end{center}
\caption{Dependence of the effective mass on temperature for $T_c/M=1/100$ (dashed/red) and $T_c/M=10$ (solid/blue).}
\end{figure}

\begin{figure}[!htbp]
\begin{center}
\includegraphics[width=0.7\textwidth]{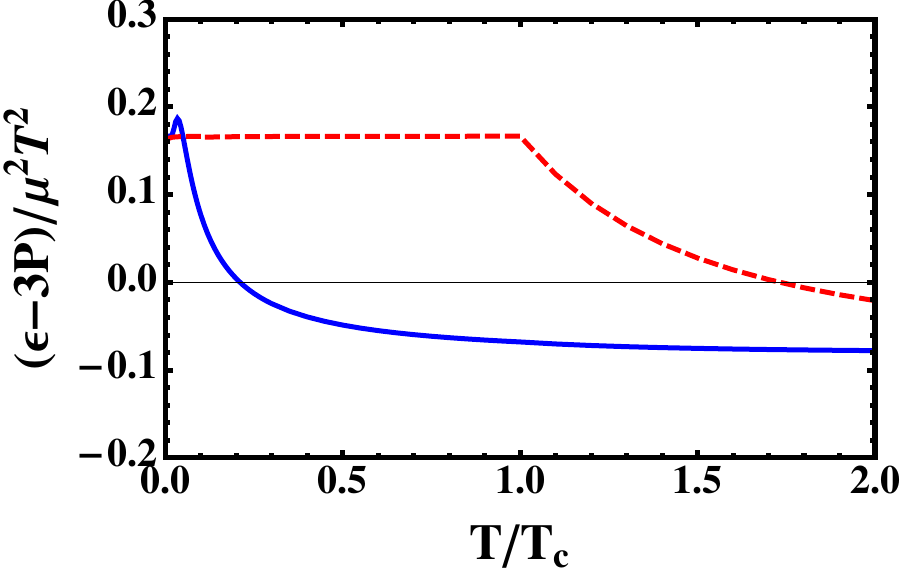}
\end{center}
\caption{The interaction measure as a function of temperature for $T_c/M=1/100$ (dashed/red) and $T_c/M=10$ (solid/blue).}
\end{figure}

\newpage

\begin{figure}[!htbp]
\begin{center}
\includegraphics[width=0.7\textwidth]{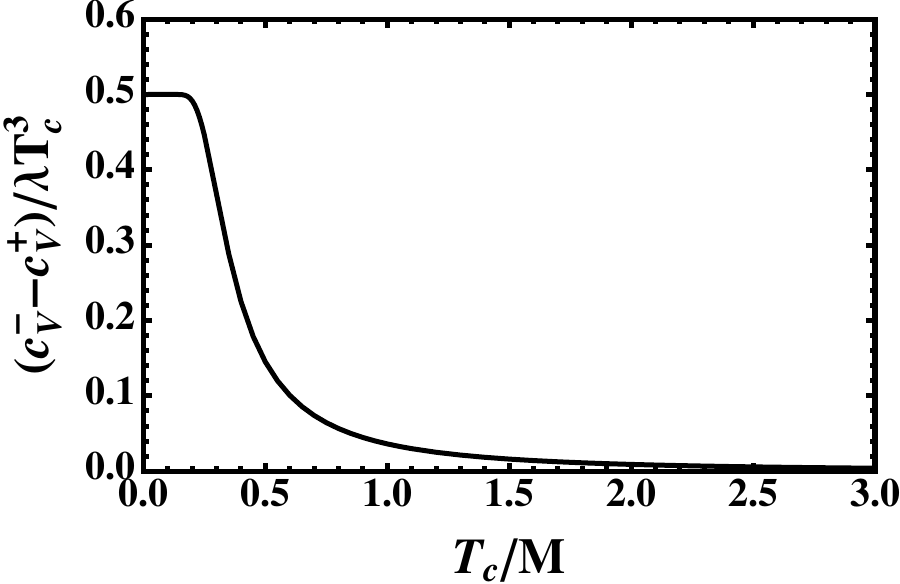}
\end{center}
\caption{Discontinuity in the heat capacity at the critical temperature.}
\end{figure}

%%%%%%%%%%%%%%%%%%%%%%%%%%%%%
\section{Effective Potential}
\setcounter{equation}{0}

For a given temperature the field $\phi$ has a stable equilibrium value $v(T)$, as discussed and computed in previous sections.  If for some reason the average deviates from its thermal value by an amount $\xi$, there will be a restoring force.  This restoring force is described by an effective potential $U(\xi)$.  In essence, this is an expansion away from equilibrium states.  It is useful in many areas of physics, including statistical physics, particle physics, and cosmology.  In this section we compute the first few terms in the expansion at the 1-loop order.

Let us define
\be
\phi=\bar\phi +\phi_f
\ee
where $\bar\phi$ is a constant field and $\phi_f$ is the fluctuation around it whose average value is zero.  Let us further write
\be
\bar\phi = v(T)+\xi \, .
\ee
Here $v(T)$ is the equilibrium value of the condensate at temperature $T$ which was previously determined.  The $\xi$ is a space and time independent (or slowly varying on all natural length and time scales) deviation from the equilibrium value that we take as a parameter to be varied at will.  We will calculate deviations from the thermodynamic potential at each temperature as a function of $\xi$; this is the effective potential (commonly referred to as the effective action in particle physics).  The Lagrangian can be written as
\be
{\cal L} = {\cal L}_{\rm quad} + {\cal L}_{\rm int} + {\cal L}_{\rm ct} - V_{\rm cl}(v) - U_{\rm cl}(\xi) - V_\xi(\phi_f)
\ee
where
\ba
{\cal L}_{\rm quad} &=& \thalf \phi_f \left[ {\rm e}^{-\Box/M^2} \left( \Box + \mu^2 \right) -12\lambda v^2 - \delta m^2 \right] \phi_f
\nonumber \\
{\cal L}_{\rm int} &=& -4\lambda v \phi_f^3 - \lambda \phi_f^4
\nonumber \\
{\cal L}_{\rm ct} &=& \thalf \left( \delta m^2 + \gamma \right) \phi_f^2
\nonumber \\
V_{\rm cl}(v) &=& - \thalf \left( \mu^2 + \gamma \right) v^2 + \lambda v^4
\nonumber \\
U_{\rm cl}(\xi) &=& \left( 4\lambda v^2 -\mu^2 -\gamma \right) v\xi + \thalf \left(12\lambda v^2 - \mu^2-\gamma\right) \xi^2 +4\lambda v \xi^3 + \lambda \xi^4
\nonumber \\
V_\xi(\phi_f) &=& 6\lambda \left( 2v\xi+\xi^2 \right) \phi_f^2 + 4\lambda \xi \phi_f^3 \, .
\ea
The first four terms in the Lagrangian were already introduced and used previously.  The last two terms vanish when $\xi=0$.  The $U_{\rm cl}(\xi)$ is the contribution to the classical potential from the $\xi$ field.  The $V_\xi(\phi_f)$ gives the interaction between the deviational field $\xi$ and the quantum field $\phi_f$; it will be used to determine the effective potential at a given temperature.

The Feynman rules corresponding to the above action are the same as before but with additional terms arising from the presence of $\xi$.
The thermodynamic potential is now
\be
\Omega = V_{\rm cl}(v) + U_{\rm cl}(\xi) - \frac{T}{V} \ln \left\{ \int [d\phi_f]
\exp \left( \int_0^{\beta} d\tau \int_V d^3x \left[ {\cal L}_{\rm quad}+{\cal L}_{\rm int}+{\cal L}_{\rm ct}-V_\xi \right] \right) \right\}
\ee
In a diagrammatic expansion the field $\phi_f$ is represented by a solid line while the external field $\xi$ is represented by a wavy line.  The vertices can easily be read off from the expressions above.  For example, the quartic interaction $\phi_f^4$ has the vertex $-\lambda$, the cubic interaction $\phi_f^3$ has the vertex $-4\lambda v$, and the cubic interaction $\xi \phi_f^2$ has the vertex $-12\lambda v$.

Now let us extrapolate away from the equilibrium value of the condensate so that $\xi$ is not equal to zero.  The effective potential for $\xi$ is
\be
U(\xi) = U_{\rm cl}(\xi) + U_{\rm loop}(\xi)
\ee
where
\be
U_{\rm loop}(\xi) = - \frac{T}{V} \ln \left\{ \frac{\int [d\phi_f] {\rm e}^S {\rm e}^{-\int d\tau d^3x V_\xi(\phi_f)}}{\int [d\phi_f] {\rm e}^S}\right\} \, .
\ee
Here $S$ is the action due to ${\cal L}_{\rm quad}+{\cal L}_{\rm int}+{\cal L}_{\rm ct}$.  This can be expanding in an infinite series in $\xi$.  The effective potential has the property that $U(0)=0$.

It is straightforward to compute the 1-loop contribution to $U$ to all orders in $\xi$.  Expand the exponential of $V_\xi$ to the $N_1$'th order in $\xi \phi_f^2$ and to the $N_2$'th order in $\xi^2 \phi_f^2$; the term $\xi \phi_f^3$ cannot contribute to the one loop order.  Expansion of the exponential gives rise to a factor $1/(N_1!N_2!)$.  Each of these $N_1 + N_2$ terms has two $\phi_f$ legs.  These must be connnected to make one and only one loop.  The ordering does not matter.  Taking into account the vertices, this leads to
\be
-\frac{(N_1+N_2-1)!}{2N_1! N_2!} (-24 \lambda v \xi)^{N_1} \left(-12 \lambda \xi^2\right)^{N_2} T \sum_n \int \frac{d^3p}{(2\pi)^3} {\cal D} ^{N_1+N_2}
\label{Usum}
\ee
Summing over all $N_1$ and $N_2$ gives  $U_{\rm 1-loop}(\xi)$ with the obvious requirement that $N_1+N_2 > 0$.  When summed with $U_{\rm cl}(\xi)$ the term linear in $\xi$ should vanish, otherwise we would not be at the extremum of $\Omega$.  This term is
\bd
\left(4\lambda v^2 -\mu^2- \gamma \right)v\xi + 12 \lambda v \xi  T \sum_n \int \frac{d^3p}{(2\pi)^3} {\cal D}
\ed
which does vanish on account of the equation satisfied by $v$ at one loop order.

Let us examine $U_{\rm loop}(\xi)$ when $T > T_c$.  Since $v(T)=0$ there are no three point vertices.
\be
U_{\rm loop}(\xi ,T) = 6\lambda \xi^2 T \sum_n \int \frac{d^3p}{(2\pi)^3} {\cal D}
+ \thalf T \sum_n \int \frac{d^3p}{(2\pi)^3} \sum_{N=2}^{\infty} \frac{(-1)^{N+1}}{N}
\left(12\lambda \xi^2\right)^N {\cal D}^N
\label{loopsum}
\ee
The reason for separating out the term quadratic in $\xi$ is that it naturally combines with the quadratice piece in $U_{\rm cl}$ to yield $\thalf m_{\rm eff}^2 \xi^2$.  The remaining terms can be exactly summed; they are referred to as the ring diagrams \cite{FT}.
\be
U_{\rm ring}(\xi ,T) =  \thalf T \sum_n \int \frac{d^3p}{(2\pi)^3} \left[ \ln\left( 1+12\lambda \xi^2 {\cal D}\right)
- 12\lambda \xi^2 {\cal D} \right]
\ee
The propagator is
\be
{\cal D}^{-1} = {\rm e}^{(\omega_n^2 + p^2)/M^2} \left( \omega_n^2 + p^2 - \mu^2 \right) + m_{\rm eff}^2 + \mu^2  \, .
\ee
The sum and integral in $U_{\rm ring}$ are dominated by $n=0$ and $p \rightarrow 0$, respectively, because there is no need for a UV cut-off in (\ref{loopsum}) for $N \ge 2$ .  The result is
\bd
U_{\rm ring}(\xi ,T) =  \thalf T \int \frac{d^3p}{(2\pi)^3} \left[ \ln\left( 1+\frac{12\lambda \xi^2}{p^2+ m_{\rm eff}^2} \right)
- \frac{12\lambda \xi^2}{p^2+ m_{\rm eff}^2} \right]
\ed
\be
= -\frac{T}{12\pi} \left[ \left(m_{\rm eff}^2 + 12\lambda \xi^2\right)^{3/2} - m_{\rm eff}^3 - 18 \lambda m_{\rm eff} \xi^2 \right] \, .
\ee
Thus the potential for $T > T_c$ is
\be
U(\xi,T) = \thalf  m_{\rm eff}^2(T) \xi^2 + \lambda \xi^4 + U_{\rm ring}(\xi ,T) + \cdot\cdot\cdot
\ee
One must be careful about using this expression too close to $T_c$ where fluctuations are large and where critical phenomena occur (nonanalytic critical exponents etc.).  It is only valid when $\xi^2 <m_{\rm eff}^2/12\lambda$ because otherwise the series does not converge to a logarithm.  This region shrinks to zero as $T_c$ is approached from above.

When $T < T_c$ a double series must be summed.  In this case we write
\be
U_{\rm loop}(\xi ,T) =  6\lambda (2v\xi+\xi^2) T \sum_n \int \frac{d^3p}{(2\pi)^3} {\cal D} + U_{\rm ring}(\xi ,T)
\ee
where
\bd
U_{\rm ring}(\xi ,T) = -\thalf T \sum_n \int \frac{d^3p}{(2\pi)^3} \left[ \sum_{N_1=1}^{\infty} \sum_{N_2=1}^{\infty} \frac{(N_1+N_2-1)!}{N_1! N_2!} (-x)^{N_1} (-y)^{N_2} \right.
\ed
\be
\left. + \sum_{N_1=1}^{\infty} \frac{(-x)^{N_1}}{N_1} +  \sum_{N_2=1}^{\infty} \frac{(-y)^{N_2}}{N_2} +x+y \right]
\ee
and where $x=24\lambda v \xi {\cal D}$ and $y=12\lambda\xi^2 {\cal D}$.  By using the integral representation
\be
(N_1+N_2-1)! = \int_0^{\infty} dt \, {\rm e}^{-t}\, t^{N_1+N_2-1}
\ee
the sums can be done followed by integation over $t$ with the result that
\be
U_{\rm ring}(\xi ,T) =  \thalf T \sum_n \int \frac{d^3p}{(2\pi)^3} \left[ \ln\left( 1+x+y\right) - x - y \right] \, .
\ee
As before, the sum and integral  in $U_{\rm ring}$ are dominated by $n=0$ and $p \rightarrow 0$, with the result that
\be
U_{\rm ring}(\xi ,T) =  -\frac{T}{12\pi} \left[ \left(m_{\rm eff}^2 + 24\lambda v\xi+12\lambda \xi^2\right)^{3/2} - m_{\rm eff}^3 - \tthalf m_{\rm eff} \left( 24 \lambda v \xi + 12\lambda \xi^2 \right)\right] \, .
\ee
Thus the potential for $T < T_c$, including both the classical and one loop contributions, is
\be
U(\xi,T) = \thalf  m_{\rm eff}^2(T) \xi^2 + 4\lambda v \xi^3+ \lambda \xi^4 + U_{\rm ring}(\xi ,T) + \cdot\cdot\cdot
\ee
This expression makes use of the solution for $v(T)$.  Previous caution concerning the radius of convergence in $\xi$ apply here as well.

Both the $T<T_c$ and $T>T_c$ potentials have the property that
\bd
m_{\rm eff}^2(T) = \frac{\partial^2 U(\xi=0,T)}{\partial \xi^2}
\ed
which is an oft-cited relationship.  However, it is only true when the self-energy is frequency and momentum independent.  The astute reader will notice that there is a contribution of order $\xi^2$ coming from $U_{\rm ring}$ when $T < T_c$ (but not when $T > T_c$.).  This contribution is of order $\lambda^{3/2}$ and therefore is subleading in an expansion in $\lambda$.  The magnitude and sign of this contribution is exactly that arising from the one loop self-energy diagram with cubic vertices, evaluated in the zero frequency and zero momentum limit; see eq. (\ref{0limitPI}).
\be
\frac{\partial^2 U_{\rm ring}(\xi=0)}{\partial \xi^2} = \Pi_{\rm cubic}(0,0)
\ee
We have already decided to drop such subleading terms.  For further discussion on this point see \cite{ArnoldEspinosa,FT}

It should be apparent that the results of this section are independent of whether the underlying action is the string field theory or the conventional local field theory.  The differences only appear when explicit functions of $m_{\rm eff}(T)$ and $v(T)$ are used along with the relationship of $T_c$ to the paramters in the action.  For example, when $T_c \ll M$ (and dropping the ring contribution)
\be
\frac{U}{\mu^2T_c^2} = \left\{ \begin{array}{ll}
{\displaystyle \frac{1}{2} \frac{m_{\rm eff}^2}{\mu^2}\left( \frac{\xi}{T_c} \right)^2 + \sqrt{2}\frac{m_{\rm eff}}{\mu} \left( \frac{\xi}{T_c} \right)^3  + \left( \frac{\xi}{T_c} \right)^4} & {\rm if} \; T \le T_c \\
{\displaystyle \frac{1}{2} \frac{m_{\rm eff}^2}{\mu^2}\left( \frac{\xi}{T_c} \right)^2 + \left( \frac{\xi}{T_c} \right)^4} & {\rm if} \; T \ge T_c \\
\end{array} \right.
\ee
and when $T_c \gg M$ (again dropping the ring contribution)
\be
\frac{U}{\mu^2T_c^2} = \left\{ \begin{array}{ll}
{\displaystyle \frac{1}{2} \frac{m_{\rm eff}^2}{\mu^2}\left( \frac{\xi}{T_c} \right)^2 + \sqrt{\frac{2\pi\sqrt{\pi}}{3} \frac{T_c}{M}} \frac{m_{\rm eff}}{\mu} \left( \frac{\xi}{T_c} \right)^3  + \frac{\pi\sqrt{\pi}}{3} \frac{T_c}{M}\left( \frac{\xi}{T_c} \right)^4} & {\rm if} \; T \le T_c \\
{\displaystyle \frac{1}{2} \frac{m_{\rm eff}^2}{\mu^2}\left( \frac{\xi}{T_c} \right)^2 + \frac{\pi\sqrt{\pi}}{3} \frac{T_c}{M}\left( \frac{\xi}{T_c} \right)^4} & {\rm if} \; T \ge T_c \\
\end{array} \right.
\ee

As examples, we show the effective potential (without the ring contribution) in Figs. 8 and 9 for $T_c/M=1/100$ and $T_c/M=10$, respectively, for $T$ below, at, and above the critical temperature.

\begin{figure}[!htbp]
\begin{center}
\includegraphics[width=0.7\textwidth]{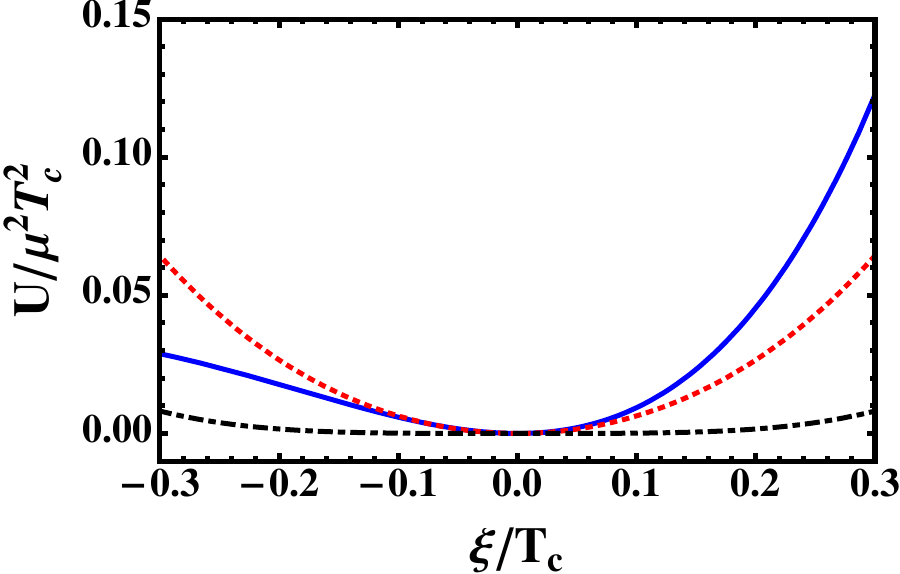}
\end{center}
\caption{Effective potential for $T_c/M=1/100$ at $T/T_c$=0.5 (solid/blue), 1.0 (dash-dotted/black), and 1.5 (dashed/red).}
\end{figure}

\newpage

\begin{figure}[!htbp]
\begin{center}
\includegraphics[width=0.7\textwidth]{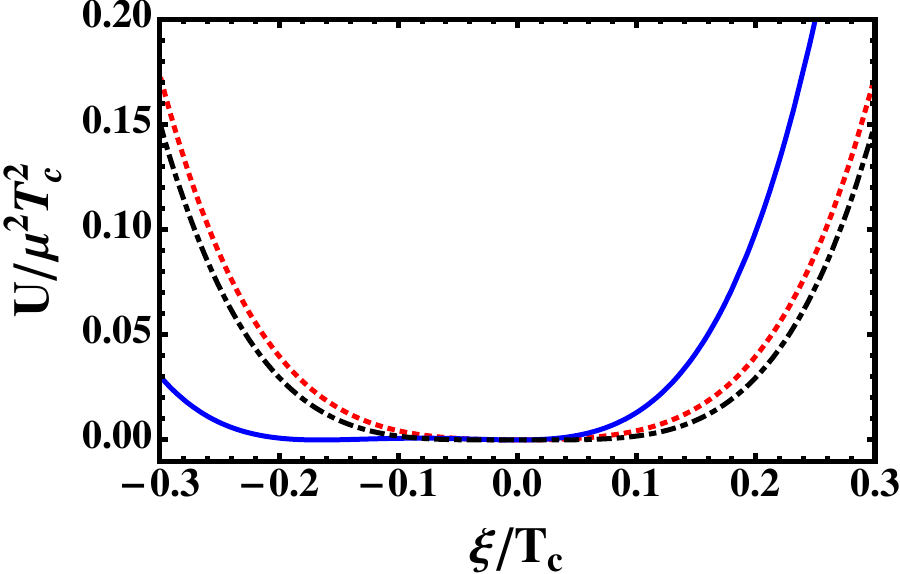}
\end{center}
\caption{Effective potential for $T_c/M=10$ at $T/T_c$=0.5 (solid/blue), 1.0 (dash-dotted/black), and 1.5 (dashed/red).}
\end{figure}

%%%%%%%%%%%%%%%%%%%%%%%%%%%%%

\section{Conclusion}
\setcounter{equation}{0}

We investigated the thermodynamic properties of a  tachyon with a string field theory motivated nonlocal action.  We first studied the phase transition in the high temperature limit to 1-loop order.  We introduced the mass shift $\delta m^2$ and calculated the thermodynamic potential around the temperature dependent true minimum, $v(T)$.  We found that at the 1-loop level both the mass shift, $\delta m^2$, and the minimum, $v(T)$, go to zero at the same critical temperature which strongly indicates a second order phase transition.

We then performed a more involved analysis.   Here we included a counter-term to allow us to make a comparison to the conventional $\phi^4$ theory.  We calculated the equation of state to 2-loop order.  We argued that the cubic contribution is suppressed compared to the quartic near the critical temperature.  We were able to analytically calculate the equation of state and hence the thermodynamic quantities in the limit $T,M>\mu$.  As expected, we found that at the critical temperature both the pressure and the entropy density were continuous but the heat capacity was not, signaling a second order phase transition.  The discontinuity in the heat capacity, in natural units of $\lambda T_c^3$, was found to decrease monotonically with $T_c/M$ .

Checking the consistency of our analysis, we made the comparison with ordinary local field theory by taking the limit that the string parameter $M\rightarrow\infty$.  Doing so we recovered the usual local field theory result.  We compared this with analytical approximations of our thermodynamic quantities for finite $M$.  We also calculated the interaction measure for both cases.  Aside from the $M$ dependence of the thermodynamic quantities in the string case, we found that the effective mass-squared vanished as $|T-T_c|$ compared to $|T^2 - T^2_c|$ in the conventional case.

In the last section we included the possibility of the the field being out of equilibrium at temperature $T$ by a small amount and computed the corresponding effective potential.  This allowed us to compute the finite temperature effective potential.  We found that we were able to calculate the 1-loop contribution to the effective potential at all orders in $\xi$ for temperature both above and below $T_c$. For both cases the ring contribution is only valid for $\xi^2<m_{eff}^2/12\lambda$.  It was also seen that both potentials satisfy the usual relationship $m_{\rm eff}^2(T) = \partial^2 U(\xi=0,T)/\partial \xi^2$.

We were able to calculate results that were consistent with conventional scalar field theory in the relevant limit.  In the limit investigated, $T,M>\mu$, we found that this nonlocal theory is very similar to the conventional one, but we were able to see effects from the stringy nonlocality. The formalism developed in this paper will help us explore the more challenging, but perhaps more interesting, case when $M \sim \mu$. Our calculations may also be relevant for capturing the thermal properties of the Early Universe which is relevant for some cosmological models .

This work was supported by the U.S. DOE Grant No. DE-FG02-87ER40328 at the University of Minnesota and by a LABoR grant at Loyola University.
%This work was supported by the U.S. DOE Grant No. DE-FG02-87ER40328.

%%%%%%%%%%%%%%%%%%%%%%%%%%%%

\end{document}